\documentclass[a4paper]{article} 
\usepackage[T2A]{fontenc} 
\usepackage[cp1251]{inputenc}
\usepackage{graphicx}

\begin{document}

\title{Multiagent's model of stock market with p-adic description of prices} 
\author {V.M.Zharkov \\614990, Perm,
Genkel st.4, \\Institute of Natural Sciences \\Perm State University} 

\maketitle

\begin{abstract}  A new multi-agent model of the stock market is formulated that
contains four states in which the agents may be located. Next, the
model is reformulated in the language of the functional integral
containing fluctuations of prices and quantities of cash flows. It
is shown that in the functional integral of that type description of
the prices is given not by the real numbers as is made in many
papers  but the p-adic numbers. It is shown in the following simple
examples extracted from  the proposed theory that the p-adic
description of prices gives good description of fractal behavior of
the trends. The formula is given for the p-adic mapping of prices.
Using this formula we obtain the main p-adic patterns which are the
same as the patterns of Elliott wave theory, this fact allows us to give
a rigorous mathematical proof of this theory. Our model is the only
model that gives a strict point like description of the fractal
behavior of prices. The developed approach opens the possibility to
give the formulation of p-adic technical analysis of the stock market.
\end{abstract}

\vspace{2in}


\section{Introduction}

 Stock market is the complex dynamical system which consists an individual traders as basic elements. 
The most fundamental approach for 
modeling stock is to use the agents which  simulate human 
traders. The second step is to derive the ensemble behavior of the whole market from the multi-
agent trading processes.  The main focus in the description of the dynamics
  in the stock market in the recent years  is to study
ensemble of traders, their collective behavior through the production of prices of the traded assets.
Many models was proposed. The most famous of them is the minority game, in which the traders can stay
two states the state of selling and buying. In many papers analytical and numerical study was done on the properties of
such models \cite{Zhang, Lux, a6, aa1, aa2}.

In this paper I consider the agents based approach to multifractal behavior of stock market. On the basis of the previously proposed agent-based model \cite{a8,aa6} the
new model with a fractal price behavior is built.  I study the appearance
of the p-adic description in the dynamics of prices in model. Efficient market hypothesis (EMH) is widely known since its inception by
Samuelson in 1965 \cite{aa1}. From the outset, it was criticized both for
its useless in practical activities and from theoretical side pointing to
the fact that some markets are predictable. EMH assumes that all information
is taken into account in prices. According to Fama efficient market is a
random walk such that the increment of prices satisfy the Gaussian
distribution. The most significant drawback EMH is the most complete
disregard for rare events or failures of large corrections.

EMH has far-reaching implications, which are discussed in majority of
financial economics textbooks -- investors are rational and homogeneous,
financial returns are normally distributed, standard deviation is a
meaningful measure of risk, there is a tradeoff between risk and return, and
future returns are unpredictable. It is well known that capital markets comprise of various investors with
very different investment horizons -- from algorithmically-based market
makers with the investment horizon of fractions of a second, through noise
traders with the horizon of several minutes, technical traders with the
horizons of days and weeks, and fundamental analysts with the monthly
horizons to pension funds with the horizons of several years \cite{aa2}.

In this paper I am interested in the behavior of investors at various
investment horizons as well as in scaling scheme of the market returns. My goal is to develop a model which takes into account different types of
investors and their investment horizons. Note that the fractal shown in
figure 2 different types of investors are marked with different depths of
the fractal or the type of Elliott Waves \cite{aa3}. Small investment
periods are described by the fractal segment of the low level, average
investment periods are represented by Elliott waves of type 1-2-3-4-5 and so
on. The very nature of the fractal principle dictates for which there is a
repetition of patterns in large scale in proportion. This discrete scale
invariance gives us some principle by which we can predict the future
movement of prices. Movement of future prices on large time scales should be
similar to the movement of prices on a small scale. Scale invariance of
fractals and Elliott waves give us a very simple principle of price
forecasting. We need a more general mathematical scheme that will give us
another mechanism that can be analyzed. Before we describe how to enter it we will give some
procedure to draw figure in price - time in a multidimensional space. For
example I can take the four-dimensional vector (open price, min, max, close
price) and denote its as one point. Then on a more coarse-grained graph we
have a point in four-dimensional space. To enter a three-dimensional space
is necessary to take a coarse grid containing the three-vector of the form $%
(P(t_i),P(t_{i+1}),P(t_{i+2})) $, here $P(t)$ - price as function of the
time. In a new functional integral approach proposed for the model, we find the regime with a 
deformed in­tegration measure in which the standard integral is replaced with the Jackson 
integral. We indicate the relation to a p-adic functional integral. For the  subsystems of traders orders and cash flow in the effective functional that results from the operator 
formulation of the Hubbard model \cite{Hubbard}, we find the two-parametric quantum derivative resulting 
in the appearance of the quantum $SU_{rq}(2)$ group. We establish the relation to the 
one-parametric quantum derivative and to the standard derivative.

This paper is devoted to justifying that a regime in which the studied macroscopic quantities are de­scribed by functions of a p-adic argument can arise in strongly correlated multielectron systems with strong multiagents interaction. We here justify the following scenario: p-adic numbers appear through a deformation of the integration measure depending on the deformation parameter q. The Jackson measure used when calculating macroscopic quantities appears in the functional integral in the strong correlation regime. In the case $q = 1/p$, where p is a prime, the Jackson integral becomes the p-adic integral \cite{Vladimirov}. We can therefore use functions of a p-adic argument to describe jumps in the Hubbard model.
In particular, we here continue the investigation of the new formulation of the functional integral that we proposed in \cite{Tower,ZharkovKirchanov}. Therefore, in what follows, we present a comparison with different versions of functional representations for creation-annihilation operators proposed in various papers. We show that the obtained expressions for supercoherent states result in more complex composite expressions for the operator symbols than those provided by the slave-boson and slave-roton approaches \cite{Hermele,Kim1}. The expressions for the creation-annihilation operators obtained here with various group reduction schemes in the Hubbard model first proposed in \cite{Tower}, result in all currently known formulations of the functional integral for systems with strong interaction. Moreover, as shown in \cite{ZharkovKirchanov}, the proposed formulation allows studying various cohomologies of groups and supergroups and thus provides a controlled concretization and expansion of dynamical groups and symmetries spontaneously appearing when strengthening the interaction in multiagents systems.
Our proposed representation for the effective functional and creation-annihilation operators is essen­tially nonlinear. This nonlinearity allows segregating those terms that provide quantum derivatives deter­mining generators of the quantum algebra in expressions belonging to the universal enveloping algebra. In \cite{Zharkov1}, we introduced the approximation in which radius vectors of vector fields were independent of the dynamical field coordinates. This assumption is equivalent to the approximation in which the values of the fields that correspond to the Casimir operators and to the invariants of the classical groups $SU(2)$ and $ SU(1,1)$ in our formulation are coordinate independent.
Below, we demonstrate that these combinations determine two parameters of deformation of the quan­tum Lorentz group, which is the group of four-dimensional rotations. We treat the transition to the one-parameter quantum derivative in detail and show that the subsystems of traders orders and cash flow of the Hubbard model are described by deformed versions of nonlinear sigma models. We further calculate the part of the functional integral for the Hubbard model that results from expressions containing this quantum derivative.
We realize our consideration on model contributions naturally present in the Hubbard model. We consider the limiting cases with respect to the deformation parameter and calculate the functional integral in these limit regimes.
We calculate the effective functional following from the kinetic energy of the Hubbard model using supercoherent states containing no more than the first powers of fermionic fields. We fix the dynamical fields such that the residual expressions indicate the difference between our approach and other approaches in the literature. The main distinct feature of our representation is its nonlinearity, which allows introducing a two-parameter quantum derivative into the problem. When passing to a one-parameter quantum derivative, we obtain a deformed nonlinear sigma model in the auxiliary dimension determined by the scale factor. Further, by analogy with the diagram technique for atomic X-operators \cite{ Tower}, we consider the integral of the scale factor linear in the quantum number. This integral determines the integration measure. Directly calculating the corresponding series, we show that in the limit case as the deformation parameter tends to zero, this series converges to the Jackson integral.
Obtaining deformations of the integration measure when varying the deformation parameter is our main result. In the limit case of small $q$, the standard integral becomes the Jackson integral.

\section{Mathematical model of agent based stock market}

This chapter is central to the search for the model of the stock market, showing fractal behavior. Below we show that we have formulated a model which can be reformulated in such a way that the effective functional gives us a p-adic description. Below we repeat parts of formulas, by presenting the material in a form that adopted in solid state physics. We briefly describe the general scheme for constructing the effective functional for the Hubbard model proposed in \cite{Zharkov1} and recently further developed, for example, in \cite{ZharkovKirchanov}. We begin with the Hubbard model written in terms of the standard creation-annihilation operators. We will take as a model the stock market model, written in terms of creation - destruction operators. This type of model is the standard representation for the Hubbard model \cite{Hubbard} in the weak interaction limit.

\begin{figure}[ht] 
\centering
\includegraphics*[width= 200pt, height=121pt]{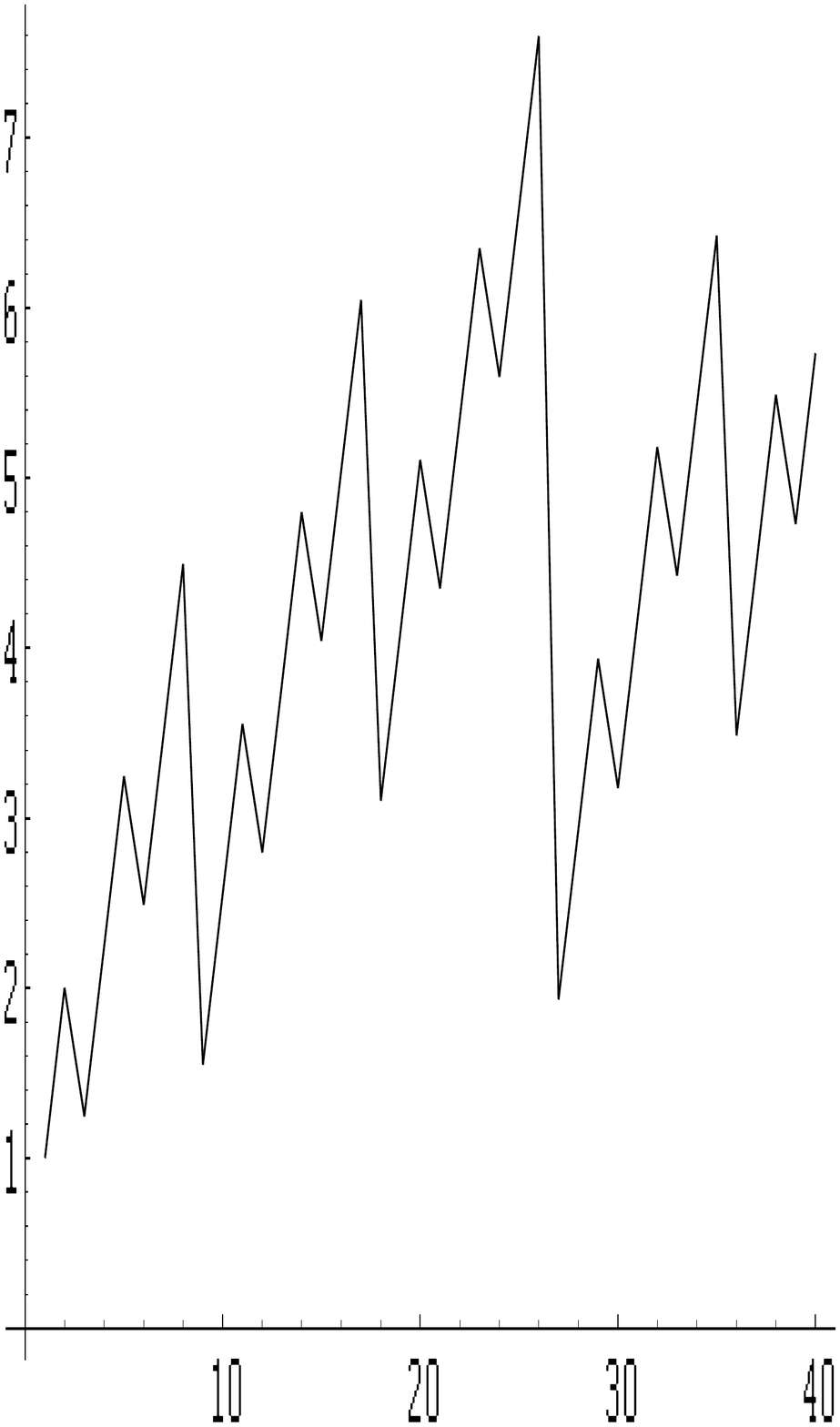}
\caption{ A subcritical wave for b < 1.
}
\label{onecolumnfigure}
\end{figure}

\bigskip

\begin{equation} H=-W\sum_{ij\sigma }\alpha _{\sigma ,i}^{+}\alpha _{\sigma ,j}+U\sum_{i,\sigma \
}n_{\sigma ,i}n_{-\sigma ,i}-\mu \sum_{\sigma ,i}n_{\sigma ,i}, \label{h1} \end{equation}

where $\alpha_{\sigma ,i}^{+},\alpha_{\sigma ,j}$  are the creation and annihilation operators,  $n_{\sigma
,i}$ is the electron density operator, $W,U,\mu $  are the respective width of the conductivity band, the one-site repulsion of two electrons, and the chemical potential, and $ \sigma $  determines the spin value. We sum over the indices $i$ and $j$ labeling the lattice sites. For instance, for a one-dimensional lattice, we have  $—N/2 < i < N/2$, where $N$ is the total number of atoms. For electrons, we sum over the indices $ \sigma=\pm 1/2 $. In the weak correlation regime with a small on-site repulsion, we take the zeroth approximation for the kinetic energy and treat the repulsion perturbatively. In the "atomic" approach with strong repulsion, we take a one-site contribution of the form
$$U n_{\sigma ,i}n_{-\sigma ,i}-\mu n_{\sigma ,i}.$$

as the zeroth approximation.
In what follows, we omit the index $i$ in order to concentrate on the internal structure of the appearing fiber bundle. This term is diagonal in the "atomic" basis and has the eigenfunctions and eigenvalues $\epsilon$:
$$|0\succ ;|+\succ =\alpha_{\uparrow }^{+}|0\succ; |-\succ =\alpha_{\downarrow }^{+}|0\succ;|2\succ
=\alpha_{\uparrow }^{+}\alpha_{\downarrow }^{+}|0\succ , $$
$$ \epsilon_0=0, \epsilon_{+}=-\mu, \epsilon_{-}=-\mu, \epsilon_2=U-2 \mu .$$

We give a mathematical formulation of the market model consisting of a set
of traders. The market consists of many traders marked the index $i $. Each trader may be in the following states: $|0\succ ;$-state in
which a trader has no money nor stock, $\mid +\succ =\alpha_{\uparrow
}^{+}|0\succ;$-state in which a trader buys shares, $\mid -\succ
=\alpha_{\downarrow }^{+}|0\succ;$-

state in which the trader sells the shares,

$\mid 2\succ =\alpha_{\uparrow }^{+}\alpha_{\downarrow }^{+}|0\succ $-

state in which the trader holds shares of $N $.

General state in which the trader resides at any one time will be given to
this function:

$\psi \left( i\right) =a_{i}\mid i0\rangle +b_{i}\mid i+\rangle +c_{i}\mid
i-\rangle +d_{i}\mid i2\rangle $

here $a_{i} ,b_{i},c_{i} ,d_{i}$ - coefficients of general state. Since all
of the state in which a trader can stay listed above us, the coefficients
have the following restriction: $\left \vert \psi \right \vert
^{2}=a^{2}+b^{2}+c^{2}+d^{2}=1$.

Problems of using of this type of functions are well known.The function of
type $a_ {i}, d_i $ is symmetric in its coordinates (bosonic type), and the
functions $b_ {i}, c_ {i} $ is antisymmetric  under the permutation of
coordinates (fermion type). The general form of the functions $\mid G> $
given by (\ref{Scs}), where the exponent is the expression $\sum_ {mn} X ^
{mn} \phi_ {mn} $, where the functions $\phi_ {mn} $ are equal to components
of the following expression$(\chi_k ^ {\ast}, \vec {h}, \vec {E}). $ The
supercoherent state (\ref{Scs}) gives us the representation of a supergroup
with a set of supergenerators $X ^ {nm}. $

We introduce the operators that take a trader out of $\mid i \succ $ in the
state $\mid j \succ.$

The calculation of matrix elements for operators of general form $\alpha_
{\sigma }^{+}, \alpha _ {\sigma} $ in this basis leads to the following
matrix representation:

For example, $\alpha _ {\uparrow }^{+}$ will have the form:

\begin{equation}
\alpha _{\uparrow }^{+}=\left(%
\begin{array}{cccc}
0 & 0 & 0 & 0 \\ 
1 & 0 & 0 & 0 \\ 
0 & 0 & 0 & 0 \\ 
0 & 0 & -1 & 0%
\end{array}%
\right) =X^{+0}-X^{2-}.  \label{op}
\end{equation}

It is seen that the operators $\alpha _ {\sigma }^{+}$ contain two nonzero
matrix elements. We introduce the operators $X ^ {rs}, r, s = 0,+,-, 2 $
containing only one nonzero matrix element. Then we have the following
expansion for the operators of creation - annihilation operators through
Hubbard ones \cite{Tower}:

\begin{equation}
\alpha _{\uparrow }^{+}=X^{+0}-X^{2-},\alpha _{\downarrow
}^{+}=X^{-0}+X^{2+},\alpha _{\uparrow }=X^{0+}-X^{-2},\alpha _{\downarrow
}=X^{0-}+X^{+2}.
\end{equation}

Since the dimension of the basis we have is 4 and the creation operators are
expressed in terms of matrix of $4 * $ 4, we generally have a base of 16
operators.

We use this notation for this operator because it coincides with the $ \gamma_5 $ operator written in the chiral basis in quantum field theory (QFT). It also equals

$$ \gamma_5=(X^{00}-X^{22})^2-(X^{++}-X^{--})^2. $$

Because we have a four-dimensional basis and the creation and annihilation operators are expressed as 4x4 matrices, we have a basis comprising 16 operators in the general case. From this set we remove the unity operators and the operators of the form

$$\gamma _5=\left( \begin{array}{cccc}
  1 & 0 & 0 & 0 \\
  0 & -1 & 0 & 0 \\
  0 & 0 & -1 & 0 \\
  0 & 0 & 0 & 1
  \end{array} \right). $$

We can express the one-site Hubbard repulsion in terms this operator. Because we can express it in terms of other operators, we do not take it into account in what follows. The remaining operators on a site in the given basis can then be separated into the fermionic operators of the form
$$(X^{0+},X^{0-},X^{+0},X^{-0},X^{+2},X^{-2},X^{2+},X^{2-}),$$

and the bosonic operators of the form

$$(X^{+-},X^{-+},X^{++}-X^{--},X^{02},X^{20},X^{00}-X^{22}).$$
We omit the lattice index of these operators to indicate the possibility of obtaining generators of the global dynamical algebra. We can endow the Hubbard operators with the lattice index by taking $N$  their copies, where $N$ is the total number of lattice sites, and constituting the direct product from these copies. In terms of these operators, the Hubbard model becomes
\begin{equation} H=U\sum_{i,r}X_{i}^{rr}-W\sum_{ij\alpha \beta }X_{i}^{-\alpha }X_{j}^{\beta }
\label{HubbAtomic} \end{equation}

Since the dimension of the basis we have is 4 and the creation operators are
expressed in terms of matrix of $4 * $ 4, we generally have a base of 16
operators.

Now we need to describe the operator describing the change in the market for
buying and selling shares of the various traders.

We describe a term that can be called the kinetic energy of the market.

\begin{figure}[ht]
\centering
\includegraphics*[width=200pt, height=121pt]{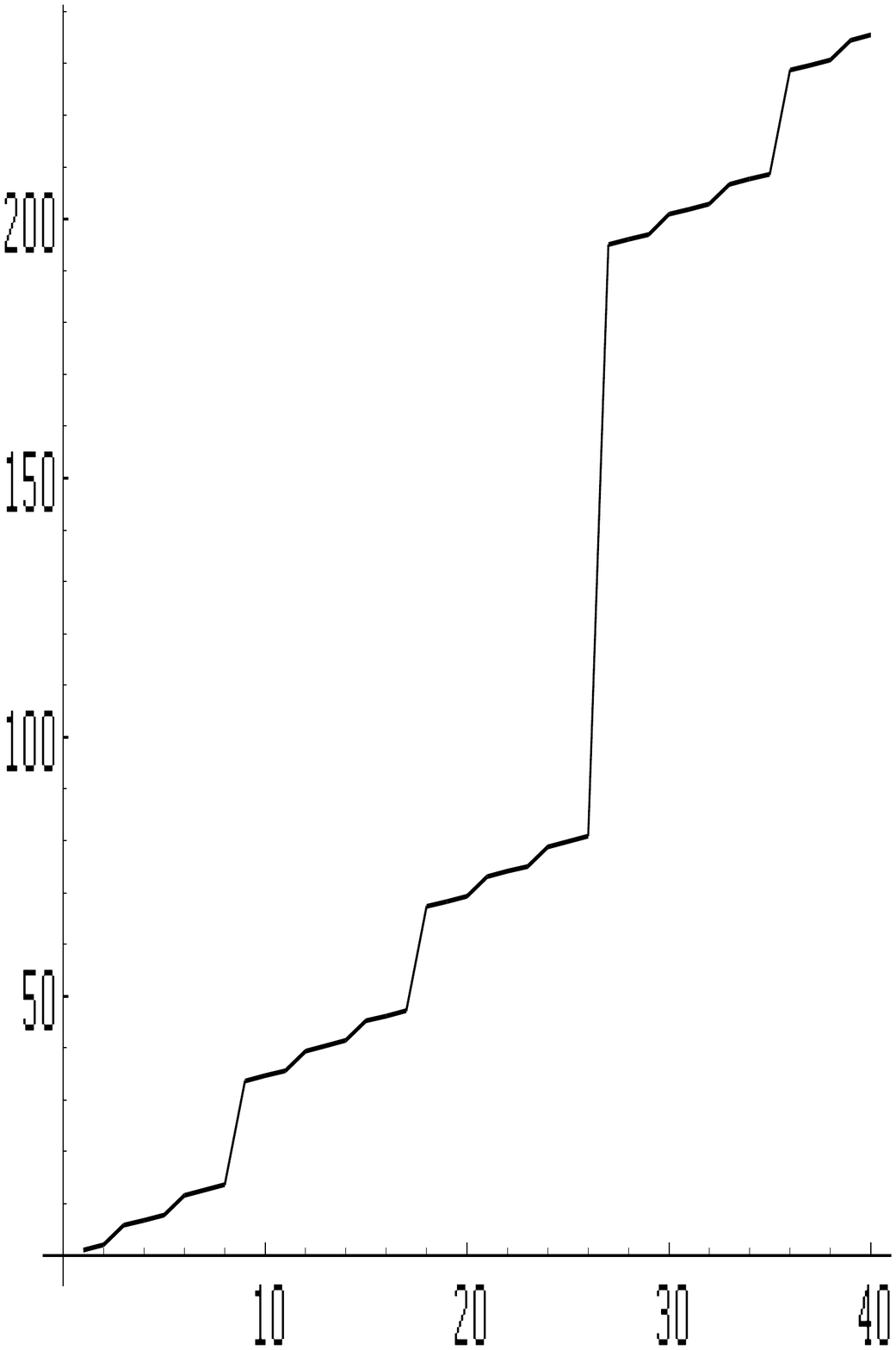}
\caption{ A supercritical wave for b > 1.
}\label{fig:name5}
\end{figure}

For example, the term $-W \sum_ {ij \alpha \beta} X_ {i} ^ {- \alpha} X_ {j}
^ {\beta} $ describes the sale of the shares of traders $i $and traders $j$
buying stocks. Other terms describe the various processes between trader $i $
and trader $j$.

The potential energy of the market is equal to:

\begin{equation}
U\sum_{i,p}X_{i}^{pp}
\end{equation}
This term describes the total volume of shares at traders. The general
solution of equation has the following form 
\begin{equation}
\mid G>=\exp \left( 
\begin{array}{cccc}
E_{z} & 0 & 0 & E^{+} \\ 
\chi _{1} & H_{z} & H^{+} & 0 \\ 
\chi _{2} & H^{-} & -H_{z} & 0 \\ 
E^{-} & -\chi _{3} & \chi _{4} & -E_{z}%
\end{array}%
\right) \mid 0>  \label{Scs}
\end{equation}

If we substitute it in formula, we obtain an expression for the matrix
elements as functions of the generalized phase (fields). We describe the
local properties of market model by this function, which is equal to the
supercoherent state.

In the exponent we have the dynamic fields that depend on the coordinates
and time.

The electric field is given by the following three-dimensional vector type:

\bigskip 
\begin{equation}
\mathbf{\vec{E}}=(E^{+}(x,y,z,t),E^{-}(x,y,zt),E^{z}(x,y,z,t)).
\end{equation}

The magnetic field has three components, depending on the spatial and
temporal coordinates

\begin{equation}
\mathbf{\vec{h}}=(h^{+}(x,y,z,t),h^{-}(x,y,z,t),h^{z}(x,y,z,t)).
\end{equation}

Fermionic fields are given grassman valued functions of coordinates and time

\begin{equation}
\chi_{k}^{\ast }(x,y,z,t),\chi_{k}(x,y,z,t),k=1,2,3,4.
\end{equation}

\begin{figure}[ht]
\centering
\includegraphics*[ width= 300pt,height=221pt]{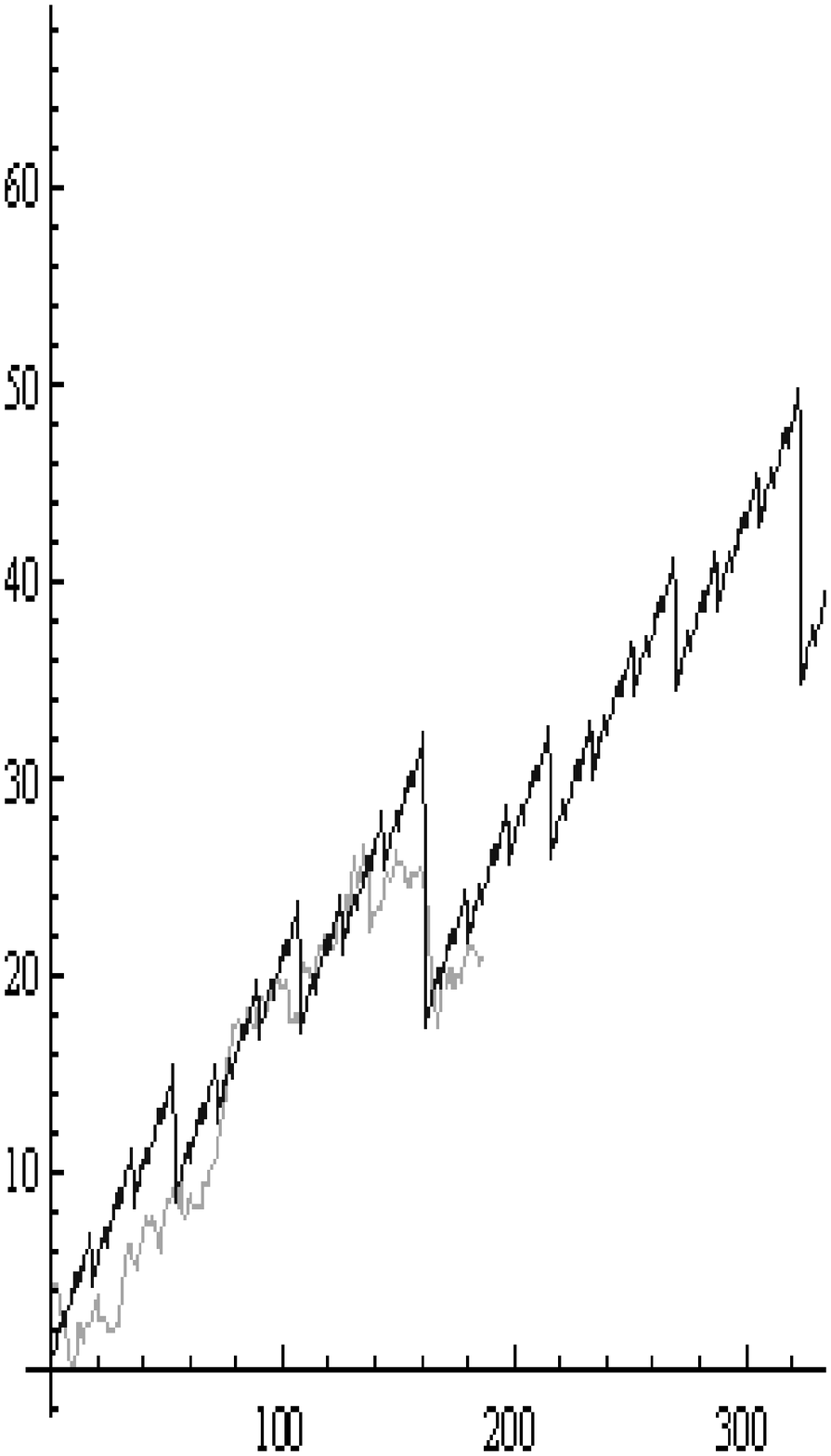}
\caption{Fractal behavior of IBM stocks.}
\label{onecolumnfigure2}
\end{figure}

Expression (\ref{Scs})  is the solution of the following problem: in the atomic basis, the site state is described by a wave function of the form
\begin{equation}
\mid G \left( i\right)> =a_{i}\mid i0\rangle +b_{i}\mid i+\rangle +c_{i}\mid i-\rangle +d_{i}\mid
i2\rangle , 
\label{func}
\end{equation}
 where $a_{i} ,b_{i},c_{i} ,d_{i} $ - are the coefficients of the atomic basis functions depending on the index  $i $.
 We have the normalizing condition for the wave function

$$ < G \mid G> ^{2}=a_i^{2}+b_i^{2}+c_i^{2}+d_i^{2}=1$$.

 Supercoherent state (\ref{Scs}) determines the supergroup representation with the set of supergenerators $ X^{nm}. $
We note that the expressions for the above supercoherent states were calculated exactly in \cite{Zhar2}. As noted in \cite{ZharkovKirchanov}, using computers and computer programs for working with supermatrices and superpoles, we can exactly calculate the functional itself. It is clear from \cite{Zhar2} that the main distinction between our approach and the approaches in other papers is the nonlinear form of the supercoherent state expression. Be­low, we show that all the representations proposed in the literature for symbols of the creation-annihilation operators follow from our representation when the fields are fixed. We also show that it is the essentially nonlinear nature of the expressions in the supercoherent state, which are functions of electric and magnetic fields, that distinguishes our representation from the previously proposed functional integral representa­tions for the Hubbard model. These nonlinear expressions contain quantum derivatives and, as a result, quantum groups. We use an absolutely different approach in this paper, and the discovery of quantum structures in the Hubbard model is therefore explainable.

The capital flow to market is given by the three-dimensional vector $\vec E $%
. The traders activity is described by three components of $\vec h $,
depending on the spatial and temporal coordinates. Fermionic fields are
given by odd grassmann's valued functions $\chi_j $ of the coordinates and
time.

Next we use a functional formulation of many-electron systems, proposed in
work \cite{a8,aa6}. According to this approach, the evolution operator
between initial and final states is given by the following functional
integral with action, expressed through effective functional, which is
calculated (\ref{HubbAtomic}) using supercoherent state.

\begin{equation}
<G_{f}|e^{-iH(t_{f}-t_{i})}|G_{i}>=\int_{|G_{i}>}^{|G_{f}>}D(G,G^{\ast
})e^{-iS[G,G^{\ast }]};  \label{fi}
\end{equation}

here the action is given by following expression: 
\begin{equation}
S[G,G^{\ast }]=\int_{t_{i}}^{t_{f}}dt\int_{V}d^3r\frac{<G(r,t)|i\frac{%
\partial }{\partial t}-H|G(r,t>}{<G(r,t|G(r,t>}  \label{lag}
\end{equation}

Measure of functional integration is equal to: 
\begin{equation}
D(G,G^{\ast }) = \prod_{t_{i}<t<t_{f}} \prod_{r} dG(r,t)^{\ast
}dG(r,t)<G(r,t)|G(r,t)> 
\end{equation}%
.

\begin{figure}[ht]
\centering
\includegraphics*[width= 300pt,height=221pt]{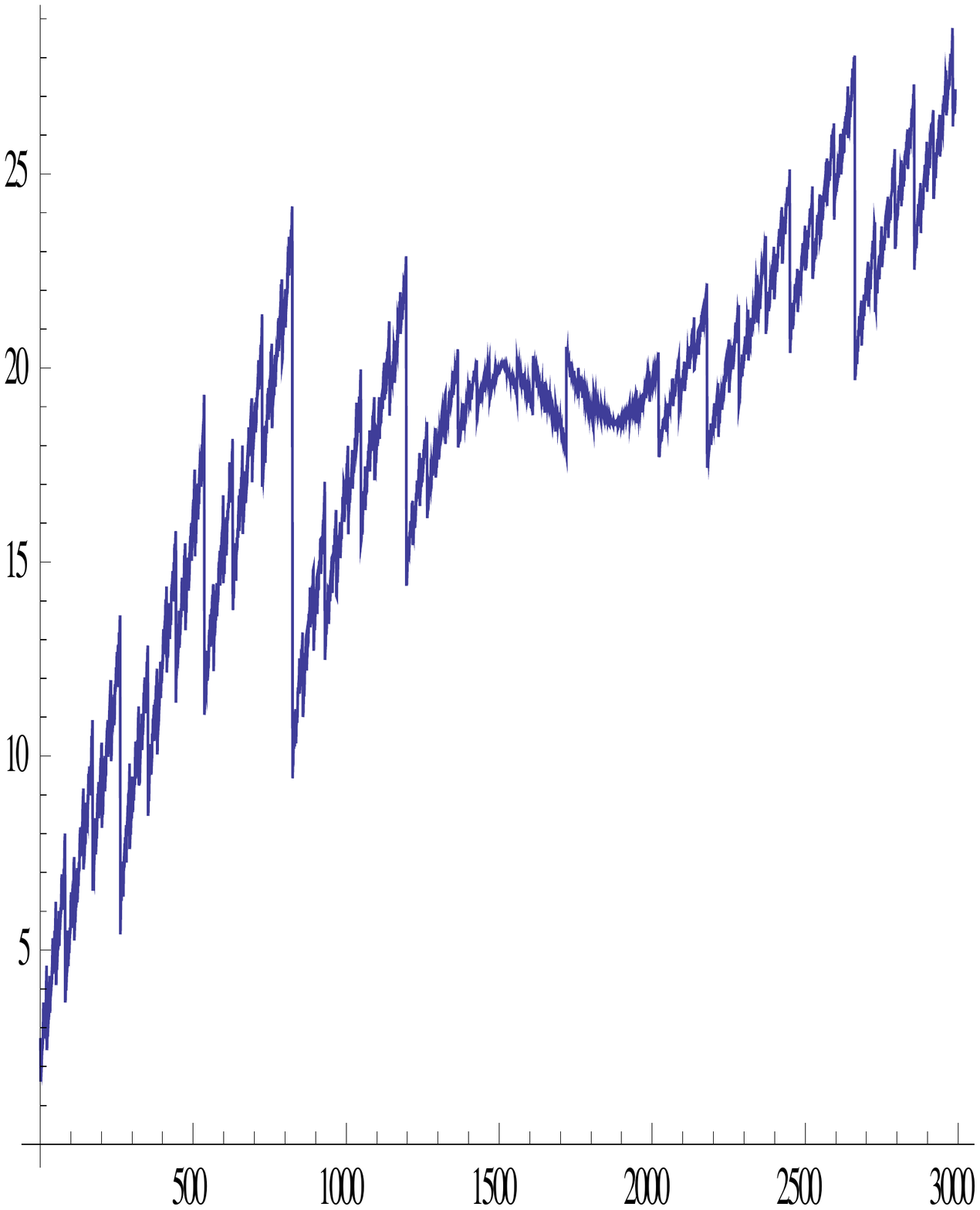}
\caption{Nonstandard correction of Elliott impulse}
\label{onecolumnfigure4}
\end{figure}


\hspace{1.5 em} We give the reasons for using supercoherent
states in the Hubbard model.

1) The first reason - is noted in numerous studies that there exist some
kind of band-atomic particle duality in strongly correlated systems
which, in regard to this matter is as follows. There are two
alternative descriptions of the same system - one offers
start of free gas and explore the Coulomb
interaction as a perturbation, and the second is based on the limit of
localized electrons and treats hopping as a perturbation.
As far as we know, the only description that can
interpolate between these two approaches is the functional approach
based on supercoherent states.

2) The second reason is the lack of clarity with dynamical symmetry inherent to
systems with strong interaction. As a consequence we have the
uncertainty of the type of functional integral. In such circumstances,
use of a common group ideology and supercoherent state is a reasonable and
promising.

\subsection{Definition of supercoherent state}


\hspace{1.5 em} Thus, the state of the system is described by the
following wave function, which depends on the coordinates $ x, y, z $ and
time $ t $ .
\begin{equation}
\label{cs} \mid G>=exp\left[
\begin{array}{cccc}
E_z & 0 & 0 & E^{+} \\
\chi _1^{*} & h_z & h^{+} & 0 \\
\chi _2^{*} & h^{-}& -h_z & 0 \\
E^{-} & -\chi _3^{*} & \chi _4^{*} & -E_z
\end{array}
\right] \mid 0>,
\end{equation}

$ \mid G> $ acts on the space $ (\mid 0>, \mid +>, \mid ->, \mid2>) $ is a basis of  "atomic" functions on which a coherent state is defined.

$$
\mid G>=\left(
\begin{array}{c}
F\chi \\
Z(E)+2B\chi_{2}\chi_{1}
\end{array}
\right),$$

here  $F = (F_{ij}) = \left(
\begin{array}{cc}
g_1^{+} & g_1^{-} \\
g_3^{+} & g_3{-}
\end{array}
\right)$, 
and the index $ i, j = 1,2 $; $ g $ coefficients are given in
appendix
                 $ B = (0, b) + \rho (\psi) Z (\tau) E ^ {-}. $

We evaluate the expression for the supercoherent state to
second order in the Grassmann fields. Next, we will compute this expression 
in Grassmann fields to the fourth degree. In this expression,
$ h ^ {\dagger} $ is the spinor representation of the magnetic vector, and
$ Z (\tau) $ - two-component spinor, equal to
$$
Z = \left(
\begin{array}{c}
\cos (\tau) +\cos (\theta ^{\prime })\sin (\tau) \\
e^{i\varphi }\sin (\tau) \sin (\theta ^{\prime })
\end{array} \right),
$$

$tg(\tau) =\frac{E\psi }{\psi ^{\prime }};$ $\rho =\sqrt{E^2\psi
^2+\psi ^{\prime 2}}$; prime means full derivative on $\alpha$,
а $\psi =\frac{\delta (E^2K)}{\delta E^2};$ $b=\frac{\partial
K}{\partial \alpha }.$

Introduce a special function
\begin {equation}
  K = (ch (\alpha
E) / E ^ 2-ch (\alpha h) / h ^ 2) / (E ^ 2-h ^ 2),
\end {equation}
  after calculations we should put $ \alpha $ to unity.

We introduce the p, q-derivative
\begin{equation}
D_{pq}f(x)=(f(px)-f(qx))/((p-q)x).
\end{equation}

Using it and the function $ f (x) = ch (\sqrt {\alpha}) / \alpha $, we write
expression for the $ K $ in the form :
\begin{equation}
D_{E^2,h^2}f(\alpha )=(ch(\sqrt{\alpha }E)/(\alpha E^2)-ch(\sqrt{\alpha }%
h)/(\alpha h^2))/(\alpha (E^2-h^2))= K(\sqrt{\alpha })/\alpha ^2.
\end{equation}

We formulate a qualitative observation: expression for SCS can be described through $ D_ {pq} $ and the usual derivatives. It is known that
$ D_ {pq} $ defines a two-parameter deformed oscillatory
algebra. We consider several cases: a) $ E_ {i} = 0, h_ {i} = 0 $, then
$$
\mid G>=\left(
\begin{array}{c}
1 \\
\chi _1 \\
\chi _2 \\
2\chi _2\chi _1
\end{array}
\right) $$

This expression coincides with the CS for fermionic operators
b) $\chi_{i}=0,h_{i}=0,$

$\mid  G>=Z(E).$

We calculate  the symbols of vector operator $ \rho_ {1}, \rho_ {2}, \rho_ {3} $.
Note that we have chosen a representation for CS selects
 the density of the ground state system. We also may to chose
the magnetic (traders states) ground state, then we would have an expression for
spin operators, with the operator $ \rho $, if
expressions replace $ E $ on $ h $. In the future, a set of operators
$ \rho_i $ we call the pseudospin (cash flow states), and $ S_i $ a spin.
\begin{equation}
\rho _3\Rightarrow z_0^{*}z_0-z_2^{*}z_2,
\end{equation}
\begin{equation}
\rho ^{-}\Rightarrow z_2^{*}z_0,
\end{equation}
\begin{equation}
\rho ^{+}\Rightarrow z_0^{*}z_2.
\end{equation}

A set of operators $ \rho_ {1}, \rho_ {2}, \rho_ {3} $ forms a
algebra. It can be seen that the above expression for us $ \rho_i $
CS to become the SU (2) $ E = i \pi / 2,3 i \pi / 2, ... $.

In this case: $\rho _3\Rightarrow cos(\theta)$, $\rho
^{-}\Rightarrow e^{i\phi}sin(\theta),$

$\rho ^{+}\Rightarrow e^{-i\phi}sin(\theta).$

We see that $ E $ parametrizes the value of the Casimir operator and gives the
magnitude pseudospin (spin). In the range of $ 0 <E <1 $ expression
differ from CS for SU (2).

Calculating the symbols of spin operators, we have the following expression for the
magnetic CS for the group SU (2): 
$S_q^{+}=h^{+}\frac{sh(h)}{h}(ch(h)+h_z\frac{sh(h)}{h});S_q^{-}=h^{-}\frac{sh(h)}{h}(ch(h)-h_z\frac{sh(h)}{h});S_q^z=ch^2(h)+\frac{sh^2(h)}{h^2}(h_z^2-h^{+}h^{-}).$

Consider the symbols for the spin operators.

Replacing the values of vector operators on average in
insulating phase $ \alpha $, we get:

\begin{equation}
S^{+}=(1-\alpha )S_q^{+}+\alpha \chi _1^{*}\chi _2;
\end{equation}
\begin{equation}
S^{-}=(1-\alpha )S_q^{-}+\alpha \chi _2^{*}\chi _1;
\end{equation}
\begin{equation}
S^z=(1-\alpha )S_q^z+\alpha (\chi _1^{*}\chi _1-\chi _2^{*}\chi _2).
\end{equation}

This expression is very meaningful. It says that in the Hubbard model
 on the site contribution to the magnetic moment (price)
given two terms: the first gives the spin density
delocalized state (Gaussian fluctuation of price)  and satisfies the linear
$ su (2) $ algebra, the second parametrizes localized contribution (fractal price) and defined
representation of the group $ SU (2) $.  The originality of this model is the value of
localized spin - it lies in the range from 0 to 1. Total
spin is the sum of two components so that the total spin is 1/2.
The same principle can be seen in the division of
vector operators.

\subsection{Evolution operator.}

\hspace{1.5 em} Below we give the procedure for obtaining of a functional integral for
Hubbard model \cite{Zharkov1}. The functional representation is desirable
especially for systems with strong interaction and the absence of small
parameter. We use the  coherent state for those supergroups that appear in
strongly correlated systems. More specifically, our challenge will be to
search for the group structure in strongly correlated systems
which will describe their specificity. We take as a basis
the  following general group structure:

1) we unite the spatial coordinates with the times and will
consider the four-dimensional space-time manifold
as the base manifold on which a maximum possible
conformal group act in the spinor representation,

2) electron operators describing the electronic degrees of freedom,
complement the group to a supergroup, which should be
superextensions of  bosonic generators. This bosonic generators give the 4D rotation group

3) considering the spin (superspin) representation of
supergroups, we construct supercoherent state that wil gives us the essentially nonlinear representation of the wave function, describing strongly correlated metal

3) using the calculated exact supercoherent state, we obtain nonlinear functional, describing an
insulating and metallic phases.

Conventional coherent states \cite {Perelomov} for simple Lie algebras
widely used for the formulation of functional integral for
Hamiltonians, in which variables have the generators of these algebras
Lee. In the case of $ SU (2) $ and $ SU (1,1) $ group the group manifold
is a sphere or a hyperboloid and is more complex
space rather than a plane, which is the phase space for
simple harmonic oscillator.

The amplitude of the transition between two states is defined by the following
expression: $ <Z_{f}|e^{-iH(t_f-t_i)}|Z_{i}> $. We need to get
expression for the effective Lagrangian in the space of state $ |
Z> $. Evolution of a quantum system in time is defined by following
operator:
$$
      U(t,t_0) = T_{ord} exp(-i \int_{t_0}^{t} H(\tau) d \tau);
$$
if $ t-t_0 =\delta t $ isa small quantity $ \delta t << 1 $, then

$$
      U(t_0+\delta t, t_0) = 1-i\int_{t_0}^{t_0 +\delta t} H(\tau) d
      \tau.
$$

It follows that the symbol of the operator have a form
$$
       U (Z, Z ^ {*} | t_0 + \delta t, t_0) = exp (-i \int_ {t_0} ^ {t_0 + \delta t} H (Z, Z ^ {*} | \tau) d
       \tau).
$$
To find the symbol $ U (Z, Z ^ {*} | t, t_0) $ divide the interval
$ [t_0, t] $ for $ N $ sites. Consider the matrix elements of 
evolution operator $ exp (-iH (t_f-t_i)) $ $ <Z_f|$ between this states and $|Z_i> $ state.
Factorize the operator $ exp (-iH (t_f-t_i)) $ $ by inserting unit
\int d \mu (Z) | Z> <Z | = 1 $

$$
<Z_f|exp(-iH(t_f-t_i))|Z_i> = \int \prod_{k=1}^N
d\mu(Z_k)<Z_f|Z_N>
$$
$$
<Z_N|e^{-i\epsilon H}|Z_{N-1}>....<Z_{k-1}|e^{-i\epsilon
H}|Z_{k}>...<Z_1|e^{-i\epsilon H}|Z_{i}>,
$$

here $ \epsilon = \frac {t_f-t_i} {N} $. To first order in this
amount we can raise the expression for the Hamiltonian to the exponential argument

\begin{equation}
\frac{<Z_{k+1}|e^{-i\epsilon H}|Z_{k}>}{<Z_{k+1}|Z_k>} =
\frac{<Z_{k+1}|(1-i\epsilon H)|Z_{k}>}{<Z_{k+1}|Z_k>} =
e^{-i\epsilon \frac{<Z_{k+1}|H|Z_{k}>}{<Z_{k+1}|Z_k>}} +
O(\epsilon^2).
\end{equation}

We obtain the following representation:

\begin{equation}
 <Z_{f}|e^{-iH(t_f-t_i)}|Z_{i}> =
\lim_{N->\infty}\int \prod_{k=1}^N d \mu(Z_k) <Z_{k+1}|Z_k>
e^{-i\epsilon \sum_{k=1}^{N}\frac{<Z_{k+1}|
H|Z_{k}>}{<Z_{k+1}|Z_k>}},
\end{equation}

here $ | Z_0> = | Z_i>; <Z_ {N +1} | = <Z_f |. $

Define the variation $ | Z>: | \delta Z_ {k +1}> = | Z_ {k +1}> - | Z_k>. $

We have:

$$
<Z_{f}|e^{-iH(t_f-t_i)}|Z_{i}> = \lim_{N->\infty}\int \prod_{k=1}^N
[d \mu(Z_k) <Z_{k}|Z_k>]<Z_f|Z_N>
$$
$$
exp(\sum_{k=1}^{N}(Ln(1-\frac{<Z_{k}|\delta
Z_{k}>}{<Z_{k}|Z_k>})-i\epsilon \frac{<Z_{k}|
H|Z_{k-1}>}{<Z_{k}|Z_{k-1}>}).
$$

Assuming that the main contribution gives the piecewise smooth trajectories we
get:
$$
\frac {d | Z>} {dt} = \frac {| \delta Z>} {\epsilon}.
$$
Hold the first order in $ \epsilon $, we obtain

\begin{equation}
<Z_{f}|e^{-iH(t_f-t_i)}|Z_{i}> =
\int_{|Z(t_i)>=|Z_i>}^{|Z(t_f)>=|Z_f>}D(Z,Z^{*}) e^{-iS[Z,Z^{*}]};
\end{equation}
$$
 S[Z,Z^{*}] = \int_{t_i}^{t_f}dt \int_{V} d{\bf r}
(\frac{<Z(r,t|i\frac{\partial}{\partial t} -
H|Z(r,t>}{<Z(r,t|Z(r,t>}
$$
$$
-i[Ln(<Z_f|Z(t_f)>)-Ln(<Z_i|Z(t_i)>)]).
$$

Integration measure is given by the following expression:

$$
  D [Z, Z ^ {*}] =
\prod_ {t_i <t <t_f} \prod_ {r} d \mu [Z (r, t) ^ {*}, Z (r, t)] <Z (r, t) | Z (r, t) >.
$$

This expression will be for us in the future major
representation for the Hubbard model.

The expression for the effective functional $S$ governs the transition from the operator formulation in terms of the creation-annihilation operators and the Hubbard operators to the field theory formulation in terms of dynamical fields of the Bose and Fermi types. As shown in \cite{ZharkovKirchanov}, the main point of this transition is the passage to the problem formulation in terms of the superbundle determined by a representation of the local supergroup of four-dimensional space rotations. In what follows, we need only the symbols of the creation and annihilation operators determined by the expressions $<G\mid \alpha _{\sigma }^{+}\mid G>,<G\mid \alpha _{\sigma }\mid G>.$  The general expressions for supercoherent states were calculated in \cite{Zhar2}. Here, we take expressions for these states in the zeroth- and first-order approximations with respect to the Grassmann fields. We write the expressions for the coherent states and their conjugates in these approximations:

$$\mid G>=\left( \begin{array}{c} ch(E)+E_{z}\frac{sh(E)}{E} \\ a_{1}^{+}\chi _{1}+a_{2}^{+}\chi
_{2} \\ a_{1}^{-}\chi _{1}+a_{2}^{-}\chi _{2} \\ E^{-}\frac{sh(E)}{E}\end{array}\right) $$

\bigskip

$<G\mid =(ch(E)-E_{z}\frac{sh(E)}{E},(a_{1}^{+})^{\ast }\chi _{1}^{\ast }+(a_{2}^{+})^{\ast }\chi
_{2}^{\ast },(a_{1}^{-})^{\ast }\chi _{1}^{\ast }+(a_{2}^{-})^{\ast }\chi _{2}^{\ast
},-E^{+}\frac{sh(E)}{E})$

\bigskip

The action of the operators on this state leads to the expressions

\bigskip

$\alpha _{\uparrow }^{+}\mid G>=\left( \begin{array}{c} 0 \\ ch(E)+E_{z}\frac{sh(E)}{E} \\ 0 \\
-a_{1}^{-}\chi _{1}-a_{2}^{-}\chi _{2}\end{array}\right) ,\alpha _{\downarrow }^{+}\mid
G>=\left( \begin{array}{c} 0 \\ 0 \\ ch(E)+E_{z}\frac{sh(E)}{E} \\ a_{1}^{+}\chi _{1}+a_{2}^{+}\chi
_{2}\end{array}\right) ,\alpha _{\uparrow }\mid G>=\left( \begin{array}{c} a_{1}^{+}\chi
_{1}+a_{2}^{+}\chi _{2} \\ 0 \\ -E^{-}\frac{sh(E)}{E} \\ 0\end{array}\right) ,\alpha
_{\downarrow }\mid G>=\left( \begin{array}{c} a_{1}^{-}\chi _{1}+a_{2}^{-}\chi _{2} \\
E^{-}\frac{sh(E)}{E} \\ 0 \\ 0\end{array}\right). $

\bigskip

The symbols for the creation and annihilation operators calculated on these states and written in the spinor form are

$$ \left( \begin{array}{c} <G\mid \alpha _{\downarrow }^{+}\mid G> \\ <G\mid
\alpha _{\uparrow }^{+}\mid G>\end{array}\right) =E_{12}^{`}a_m\chi +\hat{E}_{11}A_m\chi ^{\ast },
$$

\begin{equation} \left( \begin{array}{c} <G\mid \alpha _{\uparrow }\mid G> \\ <G\mid \alpha
_{\downarrow }\mid G>\end{array}\right) =\hat{E}_{22}am \chi +E_{21}^{`}Am\chi ^{\ast }. \label{symbol}
\end{equation}

Here, the 2x2 matrices $\hat{E}$  and $\hat{h}_i $  are

\bigskip $E_{12}^{`}=\hat{E}_{12}\left( \begin{array}{cc} -1 & 0 \\ 0 & 1\end{array}\right)
,E_{21}^{`}=\hat{E}_{21}\left( \begin{array}{cc} -1 & 0 \\ 0 & 1\end{array}\right) ,$ $$\left(
\begin{array}{cc} \hat{E}_{11} & \hat{E}_{12} \\ \hat{E}_{21} & \hat{E}_{22}\end{array}\right)= \left( \begin{array}{cc}
ch(E)+E_z sh(e)/E & E^{+}sh(E)/E \\ E^{-} sh(E)/E & ch(E)-E_z sh(E)/E\end{array}\right)$$
$$a_m=f_3 (ln(f_3/f_2))' +f_2((ln(f_2))'+E_z)\hat{h}_2. $$ $$ \hat{h}_2= \left( \begin{array}{cc} (ln(f_2))'+h_z
& h^{+} \\ h^{-} & (ln(f_2))'-h_z\end{array}\right)$$

$$\sigma=\left( \begin{array}{cc} 0 & 1 \\ 1 & 0\end{array}\right); A_m=\sigma a_m^{*}; \chi=(\chi_1,\chi_2)^T; \chi^{*}=(\chi_1^{*},\chi^{*})^T . $$

We recall the fields over which we integrate in the functional integral: the vectors $\bf{\vec{E}}$ and $ \bf{\vec{h}}$ describing fluctuations of the respective electric and magnetic degrees of freedom and the Grassmann fields $\chi_i $.  The combinations  $ E=\sqrt{E_z^2+E^{+}E^{-}}, h=\sqrt{h_z^2+h^{+}h^{-}} $, which are invariants of the two $SU(2)$ groups, also enter the supercoherent state. We here set them to be coordinate- and time-independent constants. These constants determine two deformation parameters in the quantum group problem. We now consider the coefficients  $ f_2$ and $f_3$. Expressions for these quantities are nonlinear in the above invariants. The method for calculating them exactly was developed in \cite{Zhar2}. In the nonlinear representation of supercoherent states, we have four such coeficients:$ f, f_2,f_3,f_4$. We describe the trick used to calculate these coefficients in [17]. We multiply the invariants $ E$    and $ h$ in the formulas by the scale factor x and subsequent show that we can obtain $ f, f_2,f_3,f_4$   from $ f$  by taking derivatives with respect to $ x$. At the end of the calculations we set the parameter $x$ equal to unity. In the formulas presented above, such differentiations are marked by the prime.
Expression (\ref{symbol}) determines the representation for the creation-annihilation operators in the functional integral in terms of the three components of the vector $ \vec{h} $. The representation depending on the vector $ \vec{h} $ was proposed in \cite{Hermele}. We note that in the expressions there and also here, the local group is $SU(2) * SU(2)$.
We consider particular cases of the above composite expressions for the creation-annihilation operators:
1) If $(E^{+},E^{-},E_{z})=0,(h^{+},h^{-},h_{z})=0$ ,then: $<G\mid \alpha _{\sigma
}^{+}\mid G>=\chi _{\sigma }^{\ast },<G\mid \alpha _{\sigma }\mid G>=\chi_{\sigma} $ i.e., we obtain the holomorphic representation for the creation-annihilation operators. Substituting them in (\ref{h1}), we obtain the standard representation for ferminonic systems in terms of odd Grassmann-valued fields \cite{Popov}
\bigskip 2) If $(h^{+},h^{-},h_{z})=0,$ $(E^{+},E^{-},E_{z})\neq 0$ then expression (\ref{symbol}) generates the following formulas for the functional symbols of the creation and annihilation operators:

\begin{equation} \left( \begin{array}{c} <G\mid \alpha _{\downarrow }^{+}\mid G> \\ <G\mid \alpha
_{\uparrow }^{+}\mid G>\end{array}\right) =E_{12}^{`}f_3 B\chi +\hat{E}_{11}f_3 B^{*}\sigma\chi ^{\ast
}, \end{equation}

\begin{equation} \left( \begin{array}{c} <G\mid \alpha _{\uparrow }\mid G> \\ <G\mid \alpha
_{\downarrow }\mid G>\end{array}\right) =\hat{E}_{22}f_3 B \chi +E_{21}^{`}f_3 B^{*} \sigma\chi ^{\ast
}  \end{equation}

where $ B=(ln(f_3))' + E_z. $ Setting the expression for /$f_{3}$ equal to unity, we reduce these expressions to the slave-roton representation in \cite{Kim1}.

3) If $(h^{+},h^{-},h_{z})\neq 0,$ $(E^{+},E^{-},E_{z})= 0$ , then expression (8) generates the following formulas for the functional symbols of the creation and annihilation operators\cite{Kim1}:

\begin{equation} \left( \begin{array}{c} <G\mid \alpha _{\downarrow }^{+}\mid G> \\ <G\mid \alpha
_{\uparrow }^{+}\mid G>\end{array}\right) = \hat{E}_{11}A_m\chi ^{\ast }, \end{equation}

\begin{equation} \left( \begin{array}{c} <G\mid \alpha _{\uparrow }\mid G> \\ <G\mid \alpha
_{\downarrow }\mid G>\end{array}\right) =\hat{E}_{22}a_m \chi \end{equation}

$a_m=\hat{h}_3 $, and in the logarithmic function is $f_3$, in this formula, we have  $A_m=\sigma a_m^{*}. $  We have thus shown that each representation for the creation-annihilation operators in the Hubbard model known to us is contained in some particular case of expression ((\ref{symbol}) . We note that the main feature distinguishing our representation from those in the literature is the presence of the coefficients $f_{2}$ , $f_{3}$ и $f_4$. Expressions for these coefficients were given in \cite{Zhar2}. Below, we clarify the consequences of the presence of these coefficients.

\section{ Quantum derivatives in the Hubbard model}

We see which functional results from the operator expression for the magnetic subsystem of the Hubbard model. Substituting expressions for the creation-annihilation operators in the expression for the kinetic energy, we obtain the functional

$$ H=\sum \limits_{ij}(f_{4})^2\left( \begin{array}{c} \chi_{1}^{*} \\ \chi _{2}^{*}\end{array}%
\right)_{i}^T \left( \begin{array}{c} \chi_{1} \\ \chi _{2}\end{array}\right)_{j} $$

We note that for /$f_{4}=1$, our representation coincides with the slave-roton representation used in many papers. Our aim is to study those terms in the functionals that are absent in the cited papers. We take the unit matrix as the matrix field $ \hat{E}$, which indicates the independence on the space coordinates. We mentioned in the introduction that we want to obtain the expression for the coherent state for such $ \hat{E}$. Because $ \chi $  is a fermionic field, when passing to the momentum representation, we obtain a half-filled Fermi band over which we integrate to obtain the total number of fermions, i.e., a constant. We substitute this constant for the quadratic combination of fields $\chi _{i}^{\ast }\chi _{j}$.  As the result, we obtain the expression for the effective functional at coinciding arguments $(i=j) $ :
\begin{equation} H=(f_{4})^{2} \label{Kinetic} \end{equation}

 where the coefficient $ f_4$ is \cite{ZharkovKirchanov}: $ f_4=(E sh(E)-h sh(h))/(E^2-h^2). $
 This is precisely the factor that distinguishes our functional from the one in \cite{Hermele,Kim1}. At first glance, it is just a nonlinear coefficient. But...!
We introduce the two-parameter derivative acting on a function $ f(x)$ by the formula

$$D_{rq}f(x)=(f(rx)-f(qx))/((r-q)x)$$.

It is known \cite{Chakrabarti}, that the three operators $(D_{rq},x,x\frac{\partial }{\partial x})$ constitute the two-parameter quantum algebra \ $SU_{rq}(2)$  with the following commutation relations for the generators $(S^{+},S^{-},S_{z})$ = $(D_{rq},x,x\frac{\partial
}{\partial x})$ :

$$\lbrack S_{z},S^{+}]=S^{+},[S_{z},S^{-}]=-S^{-},S^{+}S^{-}-\frac{r%
}{q}S^{-}S^{+}=[2S_{z}]_{r,q}$$  We have the definition of the two-parameter quantum number \cite{Chakrabarti}:

$$[X]_{rq}=\frac{q^{X}-r^{-X}}{q-r^{-1}}$$

\begin{figure}[ht]
\centering
\includegraphics*[width= 300pt,height=221pt]{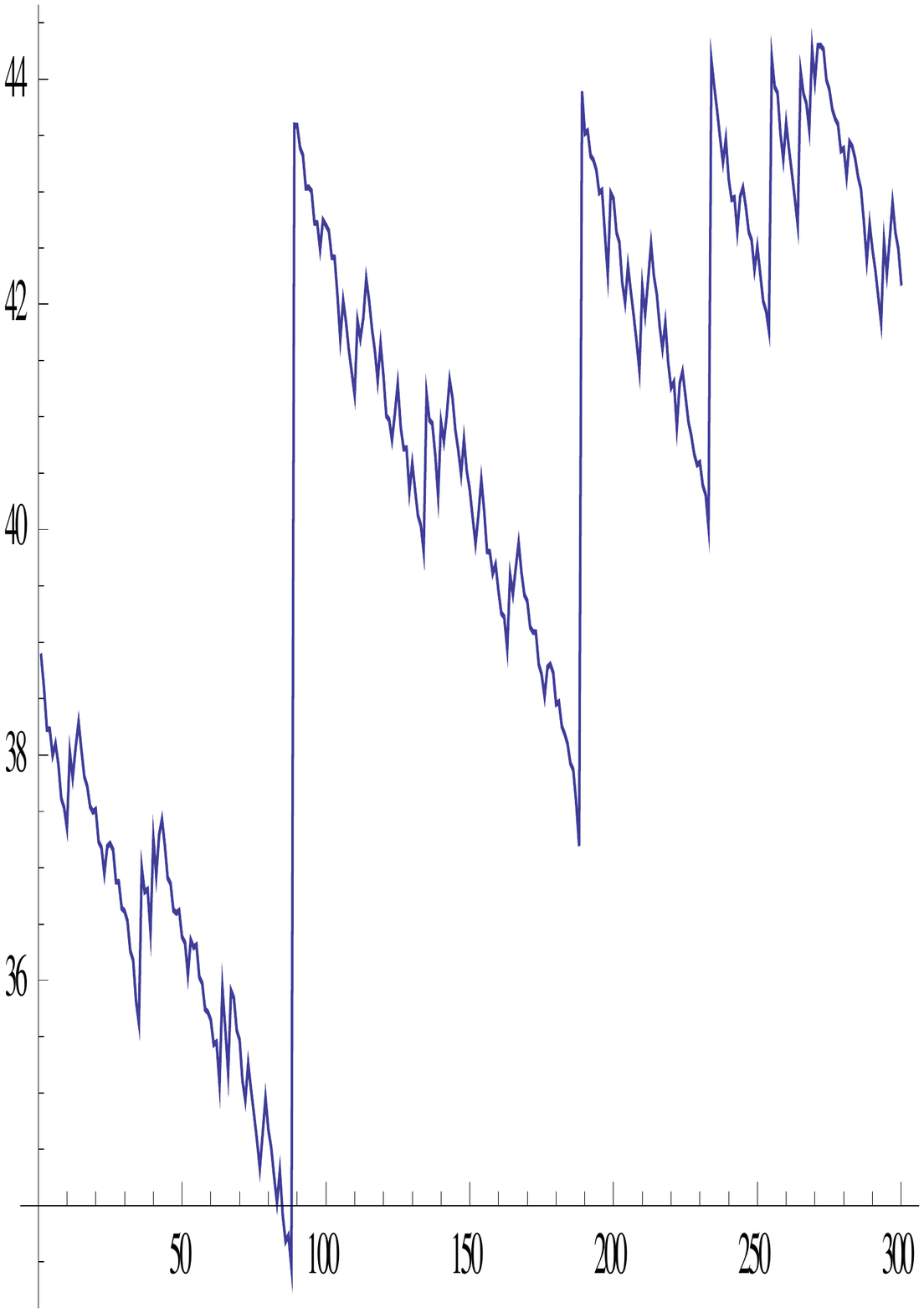}
\caption{p-adic description of triangle pattern in Elliott theory}
\label{onecolumnfigure5}
\end{figure}

Using the expression for $D_{rq}$  we can rewrite the nonlinear coefficient $f_{4}$ in the form

$$ D_{E^{2},h^{2}}(x)(\sqrt{x}sh{\sqrt{x}}/x) =(E \sqrt{x}sh(\sqrt{x}E)-h
\sqrt{x}sh(\sqrt{x}h))/(x(E^{2}-h^{2})). $$

We introduced the separate parameter $x$ into the expression for $ f_4$. The representation of this quantity in \cite{Zhar2} implies that x determines the scale factor and is the dilaton field. We now show this. The complete set of fields in expression (\ref{Scs}) for the supercoherent state and the functional integral integration measure is determined by the superalgebra whose matrix generators enter (\ref{Scs}). It does not contain the dilation generator, whose matrix representation is $\gamma_5$ in the chiral representation related to the third diagonal term in (\ref{Scs}). To take this term into account, we must calculate matrix (6) exactly with the full set of generators. Unfortunately, this problem is much harder than the problem solved in\cite{Zhar2}. We introduce the scaling factor for the local fields $\bf{\vec{E}}$ and $\bf{\vec{h}} $  such that $ \bf{\vec{E}}``->\beta \bf{\vec{E}}, \bf{\vec{h}}``->\beta \bf{\vec{h}}$  and pass to the primed fields. This factor can be a local $ x$-dependent field $ \beta(x)$,  or a global $ x$-independent field.  We can introduce a functional integration over this field, and this field then enters the integration measure. Because fields of this type (of the dilation field in QFT) had never been previously discussed in the Hubbard model framework, we below concentrate on discussing possible qualitative effects resulting from taking this field into account. In this paper, we set the scaling factor independently of the coordinate $x$. We have already shown that the quantum derivatives act on precisely this variable. We note that $ \beta(x)$,  is replaced with $x$ in formula (\ref{Kinetic}) in order to introduce a unified notation for the entity of the extended space-time coordinates. Taking the scaling factor into account means passing to the $(d+l)$-dimensional space-time in the case of the $d$-dimensional initial space-time. Substituting the derivative $D_{rq}$ in expression (\ref{Kinetic}), we obtain the functional of the form

\begin{equation} H=\sum \limits_{x}(D_{rq}(x)f(x))^{2}  \label{magnet} \end{equation}

where  $f(x)$ is given by expression in the previous formula on which the derivative acts. The difference from the nonlinear sigma model is due to specific properties of the coordinate $x$, which defines the dilaton field. We consider the limit as r$ r\rightarrow q$ or $E^{2}\rightarrow 1/h^{2}$. It is clear that $D_{rq}\rightarrow D_{q}$ in this limit and becomes the one-parameter quantum derivative acting on functions by the formula
$$D_{q}f(x)=(f(qx)-f(q^{-1}x))/((q-q^{-1})x).$$
The commutation relations for the operators ($(D_{q},x,x\frac{\partial }{\partial x})$ constitute the quantum group $SU_{q}(2)$ \cite{Pressley} with the commutation relations

$$[S_{z},S^{+}]=S^{+},[S_{z},S^{-}]=S^{-},[S^{+},S^{-}]=\frac{sh(S_{z}\ln (q))}{sh(\ln (q))}.$$
Letting $q\rightarrow 1$,  we reduce the action of the quantum derivative $D_q$ on arbitrary functions $f(x)$ to the action of the standard derivative \cite{Katz}, i.e., $D_{q}\rightarrow \frac{\partial }{\partial x}$ . Replacing $D_{q}\rightarrow \frac{\partial }{\partial x}$, we obtain the standard nonlinear sigma model $(\frac{\partial f(x)}{\partial x})^{2}$. We note that $ f$ is a solution of the normalizing constraint condition on the coherent state, which customarily defines a vector field of unit length. Such a model describes a magnetic subsystem in a functional integral approach. On the operator level, it is the Heisenberg magnet in the Hubbard model. This magnet model is defined on the dilaton field and is one-dimensional. Returning to the general effective functional in the Hubbard model, we note that we have found a quantum symmetry and the way to introduce it into strongly correlated systems. It is especially important that subsequent contractions of quantum derivatives coincide with three levels of the tower of symmetries discovered in the Hubbard model in \cite{Tower}. Formula (\ref{magnet}), implies that in multielectron systems, we must study "quantum" magnets whose variables are generators of quantum Hopf algebras, not simple Lie algebras.
We very briefly indicate the differences in the physical behaviors of such magnets. We here discuss only the behavior of the magnetic momenta and discuss more serious differences later. It is known that the
Casimir operator $S_{z}^{2}+S^{+}S^{-}$ of the group $SU(2)$ take the value: $S(S+1)$. The Casimir operator of the group $SU_{q}(2)$ is
$S^{+}S^{-}+(\frac{sh(qS_{z})}{q})^{2}$ and  it is close to zero as  $q->1$. In the physical language, this means that at the phase transition point at which a nonzero value of the parameter q appears, spin can arise smoothly with its value changing continuously from 0 to 1/2. This picture of the Mott-Hubbard-type metal-dielectric transition seems more natural compared with the phase transition scenario in which the spin instantly jumps from 0 (metal) to 1/2 (insulator of the Mott-Hubbard type). We now pass to studying a qualitative  a behavior of a functional of type (\ref{magnet}) using the functional integral.

\section{ The Jackson integral}
The above analysis of nonlinear expressions in the supercoherent state demonstrates that we can write complicated expressions in terms the action of the quantum derivative containing one or two deformation parameters. These derivatives act on functions of the variable, which was introduced in \cite{Zhar2} as a purely technical tool ensuring a uniform form of writing expressions for $ f_2, f_3, f_4$. The same variable was again used purely technically in \cite{ZharkovKirchanov} to calculate and write nonlinear expressions.
In the preceding section, we showed that by taking the scaling factor into account, we obtain a model resembling a nonlinear sigma model but with quantum derivatives instead of the standard ones. In the limit where the one-parameter quantum derivative transforms into the standard derivative, we obtain the standard nonlinear sigma model. We thus qualitatively obtained a fact that is well known in the Hubbard model: the magnetic properties of the Mott-Hubbard dielectric at large values of the Hubbard repulsion $U$ are described by the Heisenberg model with the exchange integral equal to $ W^2/U $ \cite{Hubbard}. 
Below, we use the diagram technique for spin and Hubbard operators and also our diagram technique for constructing the "tower of symmetries" [\cite{Tower}. When constructing the diagram technique for the Heisenberg model, we found the following trick useful. To the model action, which is quadratic in the spin operators, we add a term that is linear in the spin operators and proportional to $ S_z$, the so-called Zeeman contribution. We then study the quadratic terms perturbatively. Summing the diagrams responsible for the local ordering results in a nonzero contribution to the magnetic field and ensures the possibility of calculating means over the ground state when setting the previously introduced field to zero. Our scheme and expression(\ref{magnet}), imply that we encounter exactly the same situation. We cannot take (\ref{magnet}), as the leading approximation in contrast to the case where we have the standard, not quantum, derivative. We must add a linear derivative and study how it affects calculating the ground state means. Below, we show that by deforming the derivative, we do not introduce new interaction types, as could be expected, but rather deform the functional integration measure, i.e., we change the type of the functional variable itself. In addition to the standard bosonic fields described by complex-valued functions and fermionic fields described by odd Grassmann-valued functions variables, fields with a nontraditional integration measure, for instance, taking values in the field of p-adic numbers, appear in the functional integral. The base of this field is determined by the means of the bosonic electric and magnetic fields, which provide the deformation parameters for the quantum groups.
The standard scheme for studying functional integral (\ref{fi}) begins with the "zeroth" approximation ob­tained from an effective functional quadratic in either bosonic or fermionic variables, which reduces to Gaussian integrals over complex variables parameterized by momenta. We must be able to evaluate func-tionals containing quantum derivatives. For subsequent use, we briefly describe the scheme for constructing an effective functional using a holomorphic representation for creation and annihilation operators. To define an integral over trajectories completely, we must construct a functional series, which can then become a formal perturbation series. For this, we must know how to calculate integrals of the form
\bigskip \begin{equation} \int_0^{c} f(x)e^{-bx^{2}}dx=\sum \limits_{i=0}^{\infty
}\frac{f^{(n)}(0)}{n!}\int_0^{c} x^{n}e^{-bx^{2}}dx \label{series} \end{equation}
for an arbitrary function $f(x)$.
Clearly, we can exactly evaluate Gaussian integrals with power-law dependences, which allows stan­dardly constructing a formal perturbation series. Using this, for each bosonic quantum system with a Lagrangian quadratic in the fields, we can set the complex numbers $\varphi ,\varphi ^{\ast }$ into correspondence with the creation-annihilation operators $\beta _{\sigma }^{+},\beta _{\sigma },$ satisfying the standard commutation relations. Each functional integral with a bosonic Lagrangian containing interactions of fields can then be constructed as a formal se­ries in the interaction constant. To study effective functionals of type (\ref{magnet}), we must replace the exponential in expression (\ref{series}) as

$$\exp (b\varphi \varphi ^{\ast })\rightarrow \exp (bxD_{q})$$

Because we replace the pair of operators $(x,\frac{\partial }{\partial x})$ with the holomorphic coordinates $(\varphi ,\varphi ^{\ast })$, we obtain our variant of quantization if we substitute the quantum derivative  $D_{q}(x)$. for the standard one. Prom the definition of  $D_{q}$, we have

\begin{figure}[ht]
\centering
\includegraphics*[width= 300pt,height=221pt]{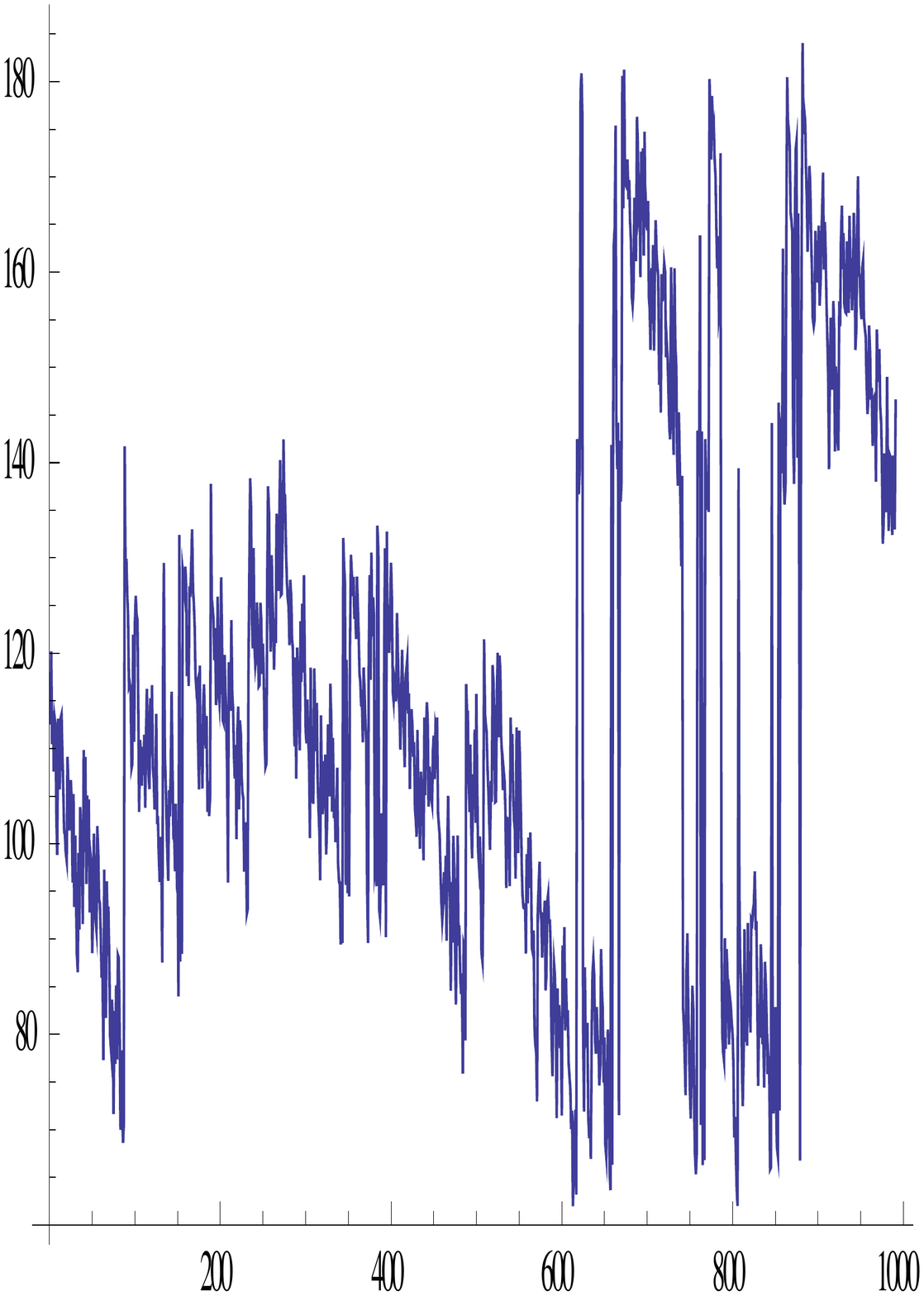}
\caption{p-adic random price signal}
\label{onecolumnfigure6}
\end{figure}

$$D_{q}\rightarrow \frac{sh(\ln (q)x)}{sh(\ln (q))x}=\frac{[x]_{q}}{x}$$

Substituting this expression in the exponent results in the expression
\begin{equation} \int_0^{c} f(x)e^{b[x]_{q}}dx  \label{Qseries} \end{equation}
where \ $[x]_{q}=\frac{q^{x}-q^{-x}}{q-q^{-1}}=\frac{sh(x\ln (q))}{sh(\ln (q))}.$
 Calculating a functional integral contain­ing general quantum derivatives $D_{rq}$, in a simpler case of the quantum derivative $D_{q}$, therefore requires calculating integrals of form (\ref{Qseries}), with arbitrary functions $f(x)$ of the variable $x$. We have thus described the scheme arising in the Hubbard model for contributions to the functional integral that contain quantum derivatives. Such integrals are always present in the general functional integral for the Hubbard model.
To calculate (\ref{Qseries}), in the general form, we expand it in the Taylor series in the derivative Dq and expand the exponential $[x]_{q}$ in the series \cite{Katz}:

\bigskip \begin{equation} \int \limits_{0}^{c}f(x)e^{b[x]_{q}}dx=\sum \limits_{m=0}^{\infty }\sum
\limits_{n=0}^{\infty }\frac{D_{q}^{m}(c)f(c)}{[m]_{q}!}\frac{b^{n}}{n!}\int
\limits_{0}^{c}(x-c)_{q}^{m}[x]_{q}^{n}dx \label{QQseries} \end{equation}

\bigskip Here $(x-c)_{q}$\bigskip $^{m}=(x-c)(x-qc)(x-q^{2}c).....(x-q^{m-1}c)$, and the quantum factorial is

$$[n]_{q}!=[n]_{q}[n-1]_{q}....[2]_{q}[1]_{q}[0]_{q}$$
The constant $c$, which determines the limit scaling in the problem, arises in (\ref{QQseries}). We represent the action of an arbitrary power of $D_{q}$ on $f(x)$ as a series,

$$D_{q}^{2n}(x)f(x)=\sum \limits_{k=-n}^{n}d_{2k}^{2n}(q)f(q^{2k}x)$$
A similar formula describes the action of an odd power of the quantum derivative on a function; the summation then ranges odd powers. Integral (\ref{QQseries}) can then be expressed as a series in $f(cq^{m})$  of the form (we present only the first terms of the series)
\bigskip $f(c)[I_{0}+\frac{d_{0}^{2}}{[2]_{q}!}I_{2}+\frac{d_{0}^{4}}{%
[4]_{q}!}I_{4}+\frac{d_{0}^{6}}{[6]_{q}!}I_{6}+...]+f(cq)[\frac{d_{1}^{1}}{%
[1]_{q}!}I_{1}+\frac{d_{1}^{3}}{[3]_{q}!}I_{3}+\frac{d_{1}^{5}}{[5]_{q}!}%
I_{5}+...]+f(cq^{-1})[\frac{d_{-1}^{1}}{[1]_{q}!}I_{1}+\frac{d_{-1}^{3}}{%
[3]_{q}!}I_{3}+\frac{d_{-1}^{5}}{[5]_{q}!}I_{5}+...]+$

$\bigskip $

$f(cq^{2})[\frac{d_{2}^{2}}{[2]_{q}!}I_{2}+\frac{d_{2}^{4}}{[4]_{q}!}I_{4}+%
\frac{d_{2}^{6}}{[6]_{q}!}I_{6}+...]+f(cq^{-2})[\frac{d_{-2}^{2}}{[2]_{q}!}%
I_{2}+\frac{d_{-2}^{4}}{[4]_{q}!}I_{4}+\frac{d_{-2}^{6}}{[6]_{q}!}I_{6}+...]$

\bigskip

Here, the quantum number \ \bigskip $\lbrack x]_{q}=\frac{q^{x}-q^{-x}}{q-q^{-1}}$ \ tend to $x$ as $q->1$. We have the representation

$$I_{n}=\int \limits_{0}^{c}(x-c)_{q}^{m}\exp (b[x]_{q})dx$$
for the quantities $I_{n}$. These integrals reduce to integrals of the form $\int \limits_{0}^{c}x^{m}\exp
(b[x]_{q})dx$  and can be evaluated as infinite series in $b$ (the detailed scheme for calculating integrals of this type and detailed expressions for the series in these integrals will be published elsewhere). As a result, we have $I_{n}\approx 1$ \ as \ $q->0.$ We present expressions for several quantum numbers:
 $\lbrack 1]_{q}=\frac{q^{1}-q^{-1}}{q-q^{-1}}=1$;

$[2]_{q}=\frac{q^{2}-q^{-2}}{q-q^{-1}}=q+q^{-1}$;

$[3]_{q}=\frac{q^{3}-q^{-3}}{q-q^{-1}}=q^{2}+1+q^{-2}$
We note that (\ref{QQseries}) reduces to (\ref{series})  as $q\rightarrow 1$. Expression (\ref{QQseries}) therefore contains a standard perturbative scheme for determining a functional integral. We consider the opposite limit $q<<1$. The expressions for quantum numbers and their factorials in this limit are \ $\ [1]_{q}\sim 1 $ , $\ [2]_{q}\sim q^{-1}$, \
$[3]_{q}\sim q^{-2}$ , \ $[4]_{q}\sim q^{-3}$ , $\ [5]_{q}\sim q^{-4}$ , \ $[n]_{q}\sim q^{-(n-1)}$  for the general term.

Expressions for the factorials for $q<<1$ \ , are: \ \bigskip $\lbrack 1]_{q}!=1$ ,
\ $\ [2]_{q}!\sim q^{-1}$ , \ \ \bigskip $\lbrack 3]_{q}!\sim q^{-3}$ , \ $\ [4]_{q}!\sim q^{-6}$ ,
\ $\ [5]_{q}!\sim q^{-10}$ , \ and: \ $[n]_{q}!\sim
\frac{1}{q^{n-1}}\frac{1}{q^{n-2}}...\frac{1}{q^{2}}\frac{1}{q}$ for the general term.
Polynomials in the small-$q$ regime are :$ p_1^1=-q;\ p_2^2=q;\ p_3^3=1;\ p_4^4=1/q^2;\ p_5^5=-1/q^5;\ p_6^6=1/q^6 .$ 
Substituting these expressions for $q<<1$  in formula (\ref{QQseries}), we obtain the series

$$\int_0^c f(x)e^{b[x]_{q}}dx \sim f(c)+f(cq)q+ f(cq^{2})q^2 + f(cq^{3})q^3 +f(cq^{4})q^4
+f(cq^{5})q^5+ .. $$
We find that expression (\ref{QQseries}) becomes the Jackson integral \cite{Katz} in the leading approximation in $q$. It is known \cite{Vladimirov}, that for $q=\frac{1}{p}$ , where $p$ is a prime, the Jackson integral becomes the p-adic integral.

\begin{figure}[ht]
\centering
\includegraphics*[width= 300pt,height=221pt]{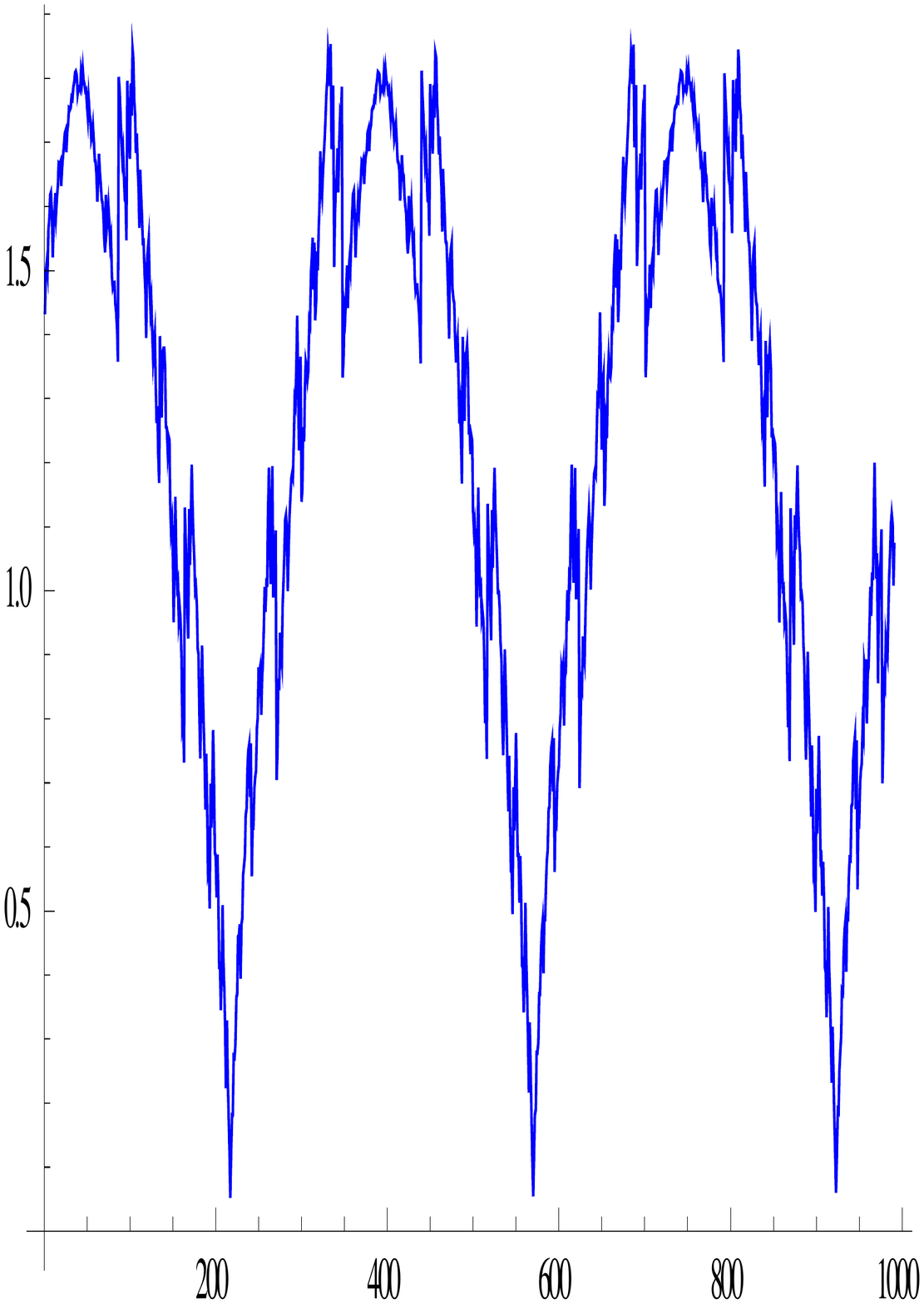}
\caption{p-adic description of sin() function}
\label{onecolumnfigure7}
\end{figure}

\section{ Elliott  Waves Theory}

We have shown the existence of the regime described by the functional integral with the Jackson measure and deformation parameter $q$ in the Hubbard model. At $q=1/p$ , in accordance with  \cite{Vladimirov}, this regime is described by the p-adic integral. To demonstrate the features of this approach, we consider some facts about p-adic numbers.

When dealing with natural numbers, if we take p to be a fixed prime number, then any positive integer can be written as a base p expansion in the form

    $\sum_{i=0}^n a_i p^i$

where the $a_i$ are integers in $\{0, p-1\}.$ Using this approach to extending  description to the rationals is to use sums of the form:

    $\pm\sum_{i=-\infty}^n a_i p^i.$

With p-adic numbers, on the other hand, we choose to extend the base p expansions in a different way. Because in the p-adic world high positive powers of p are small and high negative powers are large, we consider infinite sums of the form:

    $\sum_{i=k}^{\infty} a_i p^i$

where k is some (not necessarily positive) integer. With this approach we obtain the p-adic expansions of the p-adic numbers. Those p-adic numbers for which $ a_i = 0$ for all $ i < 0$ are also called the p-adic integers.

We briefly describe the necessary other information concerning p-adic numbers: $p$ is a prime; an arbitrary rational number $r$ admits the representation $r=p^{\nu }\frac{m}{n}$ where $n$ and $m$ are coprime to $p$. The p-adic norm of a rational number is  $\mid r\mid _{p}=p^{-\nu }$, $\mid 0\mid _{0}=0$. The field of the p-adic numbers $ Q_p$  is the completion of the field of rational numbers $Q$ with respect to the p-adic norm.

 As proposed in \cite{cc1,cc2}, we describe the prices by the function $ f_b(r)={\sum}_{0}^{N }$ $a_{k}p^{bk};\quad
a_{k}=(0,1,....,p-1);k\in Z.$  
 The coefficients  $ a_k$ are here equal to the coefficients in the p-adic number representation defined by the series r$r={\sum}^{N }_{0}a_{k}p^{k};\quad
a_{k}=(0,1,....,p-1);k\in Z.$

We take $p = 3$. The numerical modeling of this function for $b = 0.5$ and $b = 1.5$ is shown in the figures. On the ordinate, we have integers presented as series for $r$. The coefficients in $ f_b(r)$  are taken from the p-adic expansion for these numbers. The result for $ f_b(r)$  is presented on the axis OY. We set $b = 0.5$ in Fig. 1 and $b = 1.5$ in Fig. 2.

Elliott observed waves of different levels throughout the unfolding of a trend in a given 
time frame. There are certain patterns, significant price ratios and time ratios in the waves. 
The theory can be summarized in a few basic principles: 

(1) Action of a trend is followed by reaction of retracement. 

(2) There are five waves in the direction of the main trend, usually labeled as waves 1, 2, 
3, 4, 5, followed by three corrective waves, called waves a, b, c. Waves 2 and 4 are 
corrective to waves 1 and 3, and waves b and c are corrective to wave a and b 
respectively. Such a sequence of waves is also called a 5-3 move. 

(3) A 5-3 move of 8 waves completes a cycle, which then becomes 2 subdivision of the 
next higher 5-3 move. This is the key characteristic of a fractal as we know today. 

(4) The underlying 5-3 move pattern tends to remain constant, though the price range and 
time span of each wave may vary.

Diagonal

Also known as the Diagonal Triangle (leading diagonal and ends on the diagonal). Diagonal is a common 5-wave motive wave, labeled 1-2-3-4-5 that moves with a big trend. Diagonals move within two channel lines bounding drawn from wave 1 to wave 3, and from wave 2 to wave 4.

Zigzag (ZZ):

Zigzag - a 3-wave pattern, labeled ABC, generally moving against the larger trend. This is one of the most common corrective Elliott patterns.

Flat (FL)

A flat is a model of three waves labeled ABC, which generally move sideways. This corrective, counter-trend and very common Elliott pattern.

The triangle (triangle convergent (CT) and expanding triangle (ET)):

The triangle is a normal corrective pattern with 5 waves, labeled ABCDE, which move against the trend. Triangles move within two channel lines drawn from wave A to wave C, and from wave to wave B D. The triangle pattern is converging or diverging depending converge or diverge line of the channel.

In author paper \cite{aa6} it was shown that a multi-agent model which is
proposed in this paper leads to the p-adic description of price series. And
this description perfectly describes fractals. The advantage of the p-adic
description of fractals is that the fractal behavior of prices is defined
locally. We describe fractals as pointwise. This description gives an exact
match with the theory of Elliott patterns, which means rigorous mathematical
basis of this theory by p-adic mathematics. Below we give some examples of
application of p-adic functions to the price series. In fig.1 I give the
p-adic impulse pattern of Elliott theory. In fig.2 I give alternative stepwise wave. 

Below (fig.3) the graphs of IBM shares are shown. The second curve shows the
procedure a p-adic interpolation of real data. Different time scales are
used. The second curve shows the real data. It is seen that qualitatively,
the p-adic functions well describe the real data. We also see that the
p-adic functions give a good description of the patterns that appear in the
theory of Elliott waves. In this theory, the emergence of trend is
associated with the pattern, which is called a triangle. Triangle defines
the behavior of volatility in time. When the volatility tend to zero, in
this point the new trend appears. In fig.5 a triangle is shown. It can be
described if we take the number field base equal to a rational number. In
fig. 4 I show an alternative correction after an impulse Elliott wave
containing five small pulses. In fig. 6 I show that the p-adic functions
can give a signal that looks like random, right signal is made up of deep
irregular fluctuations. In fig. 7 I construct a p-adic function of the
sin(). It can be seen that the harmonic function acts as an envelope for the
fractal behavior.

\section{ Conclusion}

We have shown that the regime described by the p-adic functional integral exists in the 
functional integral for the Hubbard model.
Finally, we stress that the functional integral in the Hubbard model, as shown in this paper, 
can be successfully reformulated in terms of both the standard and the Jackson integral and 
that the latter can be reduced to the p-adic integral in a particular case. We have the 
impression, supported by the comparison with experimental prices data that in the strong correlation regime, the description of multi-electron systems is 
invariant under the choice of the number field. We can begin with either a p-adic or real 
number description. It seems plausible that this situation can be clarified if an exact 
analytic calculation of coefficients in expressions (\ref{QQseries}) would be possible. 
Because our main regimes were \ $q->1$\ and \ $q->0$, it seems plausible that we should 
study the crossover regime from the p-adic to the standard perturbation series based on 
Gaussian integration in more detail if we will be able to calculate series (\ref{QQseries})
 more exactly. This regime would clarify an interesting question: which order parameter 
appears in the system upon the introduction of the p-adic description whose distinct 
feature is the appearance of fractal behavior in physical quantities?

We have shown that the proposed model of the stock market that contains a
lot of traders leads to two regimes: 1) a p-adic regime, describing the
fractal behavior of prices \cite{aa3,aa6},  and 2) fluctuation mode
behavior of prices with a Gaussian distribution .

\end{document}